\documentclass[aps,reprint,amsmath,amssymb,floatfix,superscriptaddress]{revtex4-1}

\usepackage{physics}
\usepackage{color}
\usepackage{latexsym}
\usepackage{graphicx}
\usepackage{times,psfrag,subfigure}
\usepackage{enumerate}
\usepackage{amsmath}
\usepackage{dsfont}
\usepackage{dcolumn}
\usepackage{bm,bbm}
\usepackage{latexsym,amsmath,amssymb,bm,euscript}
\usepackage[normalem]{ulem}
\usepackage{dsfont}
\usepackage{textcomp}
\usepackage{tabularx}
\usepackage{setspace}
\usepackage{ctable}
\usepackage{sidecap}
\usepackage{placeins}
\usepackage{threeparttable}
\usepackage{multirow}
\usepackage[hidelinks]{hyperref}

\begin{document}
\title{%
Perspective: Molecular beam epitaxy of antiperovskite oxides
}
\author{H. Nakamura}
\affiliation{Department of Physics, University of Arkansas, Fayetteville, AR 72701, USA}
\author{D. Huang}
\affiliation{Max Planck Institute for Solid State Research, 70569 Stuttgart, Germany}
\author{H. Takagi}
\affiliation{Max Planck Institute for Solid State Research, 70569 Stuttgart, Germany}
\affiliation{Department of Physics, University of Tokyo, 113-0033 Tokyo, Japan}
\affiliation{Institute for Functional Matter and Quantum Technologies, University of Stuttgart, 70569 Stuttgart, Germany}

\begin{abstract}
Antiperovskites, or inverse perovskites, have recently emerged as a material class with a plethora of promising electronic properties. This perspective describes the molecular beam epitaxy (MBE) growth of oxide antiperovskites Sr$_3$PbO and Sr$_3$SnO. We show that MBE offers great potential not only in growing antiperovskites with high structural quality, but also in providing a means to seamlessly connect with advanced characterization tools, including x-ray photoelectron spectroscopy (XPS), low-energy electron diffraction (LEED), reflection high-energy electron diffraction (RHEED), and scanning tunneling microscopy (STM), to facilitate the analyses of their intrinsic properties. The initial results point toward the feasibility of atomically controlled antiperovskite growth, which could open doors to study topological and correlated electronic states in an electronic environment quite distinct from what is available in conventional complex oxides.

\end{abstract}
\maketitle

\section{Introduction}
Antiperovskites with $A_3B$O structure ($A$ = Ca, Sr, Ba, Yb, Eu; $B$ = Sn, Pb) have recently acquired considerable interest in the condensed matter physics community, in part due to the prediction of their topological electronic properties. From a chemistry perspective, the $A_3B$O antiperovskites belong to the Zintl compounds, which are characterized by a nearly complete electron transfer from the more electropositive $A$ atom to the more electronegative $B$ atom \cite{WIDERA19801805,Jansen2004,Nuss:dk5032}. In this case, the alkaline earth cation (A$^{2+}$) and the electronegative ion (B$^{4-}$) both achieve a closed-shell valence state, resulting in a narrow-gap semiconductor or semimetal that may host topological phases. In recent years, the hunt for new topological materials has opened researchers' eyes to wider materials within the Zintl phase. The Dirac semimetal Na$_3$Bi is a well known example of Zintl phase~\cite{doi:10.1126/science.1245085}. Sr$_2$Pb and Sr$_2$Sn, which were proposed to be topological insulators under strain~\cite{PhysRevB.84.165127}, are also Zintl compounds closely related to the $A_3B$O antiperovskites. Unusual electronegative Sn and Pb ions in antiperovskites have been experimentally verified in recent works~\cite{Huang_PRM_2019,PhysRevB.100.245145}. 

Three-dimensional (3D) Dirac states with a small mass gap $E_g$ were identified theoretically in antiperovskites by Kariyado \textit{et al}~\cite{Kariyado_JPSJ_2011, Kariyado_JPSJ_2012, Kariyado_PRM_2017}. Later, Hsieh \textit{et al.} gave a rigorous topological classification, showing that some of the $A_3B$O antiperovskites are topological crystalline insulators (TCIs)~\cite{Hsieh_PRB_2014}. Many theoretical studies followed, clarifying the nature of surface states~\cite{Chiu_PRB_2017}, superconductivity~\cite{Kawakami_PRX_2018}, hinge states~\cite{Fang_PRB_2020}, and surface magnetism~\cite{Arras_PRB_2021}. One of the most appealing features is that there is a large energy window in the $E(k)$ diagram ($\sim$0.4 eV in Sr$_3$PbO) in which topologically interesting states, such as the bulk 3D Dirac bands, do not overlap in energy with topologically trivial bands. The existence of such an energy window offers a unique opportunity to study topological electronic states. Furthermore, the Dirac nodes are located at the Fermi level ($E_F$) for stoichiometric compounds, which is beneficial to the study of those states by electronic transport.

Transport and thermoelectric properties of $A_3B$O antiperovskites have also been extensively studied experimentally. The observations of superconductivity in highly Sr-deficient Sr$_{3-x}$SnO~\cite{Oudah_NatComm_2016}, large thermoelectric power~\cite{okamoto_jap_2016}, quantum oscillations~\cite{Suetsugu_PRB_2018}, and large diamagnetism associated with Dirac electrons~\cite{Suetsugu_PRB_2021}, whose dispersion was also confirmed by ARPES measurements~\cite{Obata_PRB_2017}, were some of the early data obtained using bulk crystals. Molecular beam epitaxy (MBE), first developed for Sr$_3$PbO~\cite{Samal_APLM_2016}, enabled tunable carrier densities ($n$) in Sr$_3$SnO and facilitated detailed analyses of three-dimensional weak antilocalization~\cite{Nakamura_NatComm_2020}. Alternative MBE growth approaches for Sr$_3$SnO were also reported~\cite{Ma_AM_2020, Wu_APL_2021}. All the experimental evidence, including ARPES, points to the correctness of the bulk band structure predicted by theories, although critical information such as the value of $E_g$ in $A_3B$O antiperovskites is still missing. The experimental evidence of TCI surface states is also thus far lacking.

Despite these experimental and theoretical progress, the understanding of antiperovskites is still at its infancy. For example, we do not know what kind of impurity phases could exist in this system, especially when the stoichiometry is deliberately changed from pure compositions. This complicates the interpretation of superconductivity observed in Sr$_{3-x}$SnO, where secondary phases were also found~\cite{Oudah_NatComm_2016}. Another question is associated with the interface or surface of antiperovskites. Along the [001] direction of antiperovskites is a sequence of positively ($+2$) and negatively ($-2$) charged layers ($BA$ and O$A_2$, respectively), which should trigger electronic or atomic reconstructions at the surface or interface, as per polar discontinuity~\cite{Nakagawa_NatMat_2006}. The physics of reconstructions is currently poorly understood for antiperovskites. On the other hand, because of the close match in lattice constants among antiperovskites, fabricating perfect interfaces of topological materials seems feasible. Emerging phases at interfaces is another agenda that deserves continued effort. A recent discovery of interface superconductivity at the KTaO$_3$(111) surface, at a much higher $T_c$ than that observed on the (001) surface, reaffirms that even the interfaces of conventional oxides are still rich in unexpected electronic phenomena~\cite{Liu_Science_2021}. Antiperovskites hold strong promise for the discovery of novel electronic phenomena at the interface, especially once aspects of material growth are better understood.

\begin{figure}[!h]
\includegraphics[width=8cm,clip]{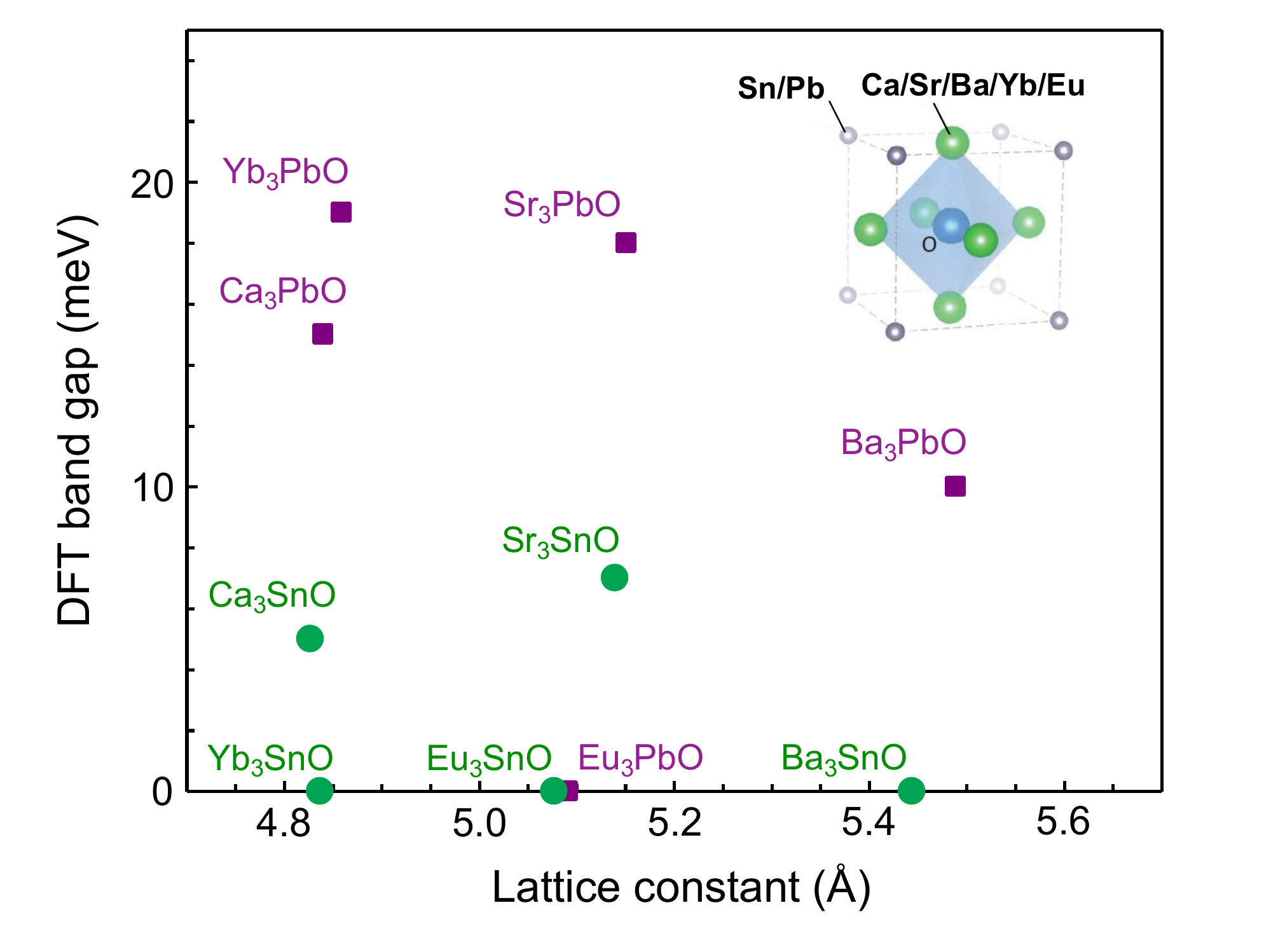}
\caption{%
Band gap vs. lattice constant of selected $A_3B$O antiperovskite materials ($A$ = Ca, Sr, Ba, Eu, Yb; $B$ = Pb, Sn). All the lattice constants come from X-ray diffraction data~\cite{Jansen2004,Nuss:dk5032}, while the band gaps are computed from first-principles DFT calculations~\cite{Chiu_PRB_2017, Pertsova_PRB_2019}. For Ba compounds, lattice constants for the room temperature cubic phase are shown.
}
\label{fig:1}
\end{figure}

The structure-property relationship of different antiperovskites within the $A_3B$O family may be understood by focusing on two key parameters, bandgap ($E_g$) and lattice constant. As shown in Fig.~\ref{fig:1}, the lattice constants of antiperovskites are largely determined by the element $A$. For instance, $A_3B$O antiperovskites with $A$ = Ca and Yb show similar lattice constants, so do the ones with $A$ = Sr and Eu, largely independent of whether the $B$ element is Pb or Sn. Next, antiperovskites with Pb show universally larger band gaps compared to their Sn counterparts. This is because the opening of the small band gap ($\sim$10 meV) is controlled by the strength of spin-orbit coupling. An exception to this simple rule is the antiperovskites with Eu, where $E_g$ disappears both for the Pb and Sn compounds, suggesting the influence of $f$ electron states~\cite{Pertsova_PRB_2019}. The valence and conduction bands consist of the $p$ orbitals of Sn/Pb and the $d$ orbitals of alkaline earth elements, respectively, and the formation of Dirac bands is driven by a band inversion at the $\Gamma$ point. This generates six identical Dirac nodes along equivalent $\Gamma$--$X$ lines. Since the Dirac nodes are located at $E_F$, the small gaps shown in Fig.~\ref{fig:1} are the ones at Dirac crossings, making the Dirac electrons massive.  

In this Perspective, we provide insights into the MBE growth of Sr$_3$PbO and Sr$_3$SnO antiperovskites. We show that MBE is one of the ideal techniques to study the intrinsic properties of as-grown antiperovskites, which are air-sensitive, as is the case with many other Zintl compounds. The study of the two antiperovskites, grown and analyzed using identical methods, sheds light on finer aspects of structure-property relationships. This paper is organized as follows. In Subsection \ref{co-depo}, we discuss a global phase diagram for Sr$_3$PbO that is mapped out by co-depositing elements under vacuum with varying flux ratios. In Subsection \ref{seq}, we discuss a sequenced shutter approach, inspired by many earlier MBE works, that we believe could improve the quality of Sr$_3$PbO films, and also enabled us to quickly stabilize the Sr$_3$SnO phase. In Subsection \ref{alt}, we describe an alternative approach to the MBE of Sr$_3$SnO by using SnO$_2$ as a solid source. In Subsection \ref{trans}, we discuss \textit{ex-situ} transport measurements of the films. In Subsection \ref{surf}, we discuss various characterizations of the surface of antiperovskite films in ultra-high vacuum (UHV). We note some distinctions between Sr$_3$PbO and Sr$_3$SnO that highlight their complex chemistries, including signatures of surface reconstructions in the latter, but not in the former.
   
\section{MBE growth of $A_3B$O antiperovskites}
\subsection{Identification of Growth Window: Co-deposition}
\label{co-depo}

The lattice constants of $A_3B$O antiperovskites are $a \sim 5 \,\mathrm{\AA}$, significantly larger than the standard perovskite lattice constant of around 4 $\mathrm{\AA}$~\cite{schlom_2008_JACeram}. As noted earlier, Sr$_3$PbO ($a=5.15 \,\mathrm{\AA}$) and Sr$_3$SnO ($a=5.14 \,\mathrm{\AA}$) have almost identical lattice constants due to a common $A$ element, and can be grown on the same substrate of choice. Convenient substrates include yttria-doped ZrO$_2$ (YSZ) with $a$ = 5.14 \AA~($-$0.2\%), LiAlO$_2$ with $a=5.17 \,\mathrm{\AA}~(+0.35\%)$, LaAlO$_3$(LAO) with $a=3.79\times \sqrt{2} = 5.36 \,\mathrm{\AA}~(+4.1\%)$, and silicon with $a=5.43 \,\mathrm{\AA}~(+5.4\%)$. SrO, which is an easily grown binary oxide, also functions as a perfect buffer layer with $a=5.16 \,\mathrm{\AA}~(+0.2\%)$. Here, we consider the (001) facet of these substrates for growth, the percentage denotes the lattice mismatch with respect to Sr$_3$PbO, and a $\sqrt{2}$ factor for LAO reflects an in-plane 45$^\circ$ rotation of the film unit cell with respect to the substrate. An initial report on the MBE growth of $A_3B$O was for Sr$_3$PbO on LAO~\cite{Samal_APLM_2016}. Despite a moderate lattice mismatch of 4.1$\%$, this choice of substrate was important in the initial stages of film growth, since it allowed an easy identification of all the antiperovskite ($00l$) x-ray diffraction (XRD) peaks without overlap with the substrate peaks.

\begin{figure}[!tb]
\includegraphics[width=8cm,clip]{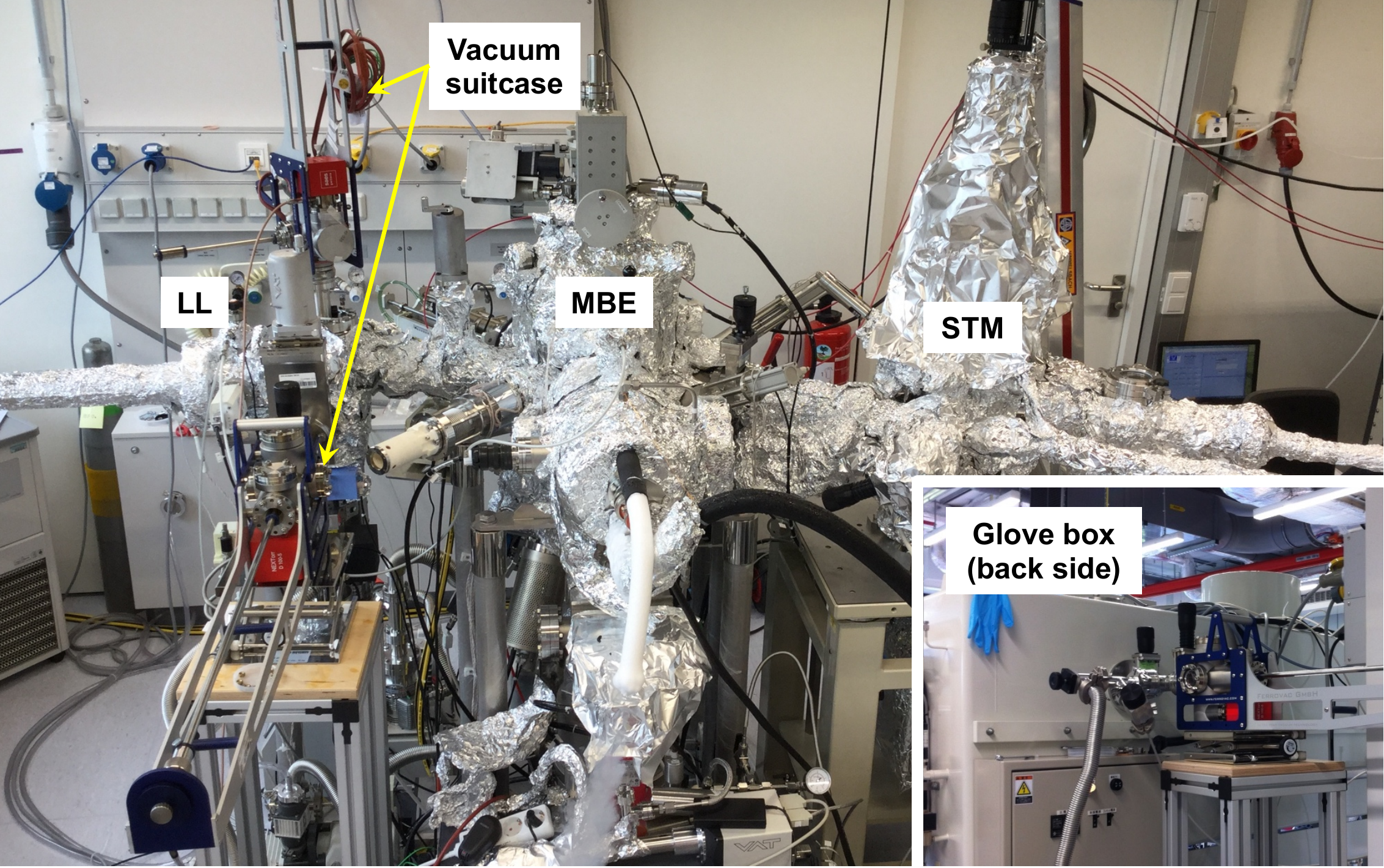}
\caption{%
Antiperovskite MBE system as developed at the Max Planck Institute for Solid State Research in Stuttgart, Germany. A room-temperature STM is integrated to enable direct transfer of grown films from the MBE. Two UHV suitcases could be mounted to the load lock (LL) chamber of the MBE, allowing transfer to a neighboring glove box (inset), as well as external facilities.
}
\label{fig:mbe}
\end{figure}

\begin{figure}[!tb]
\includegraphics[width=8cm,clip]{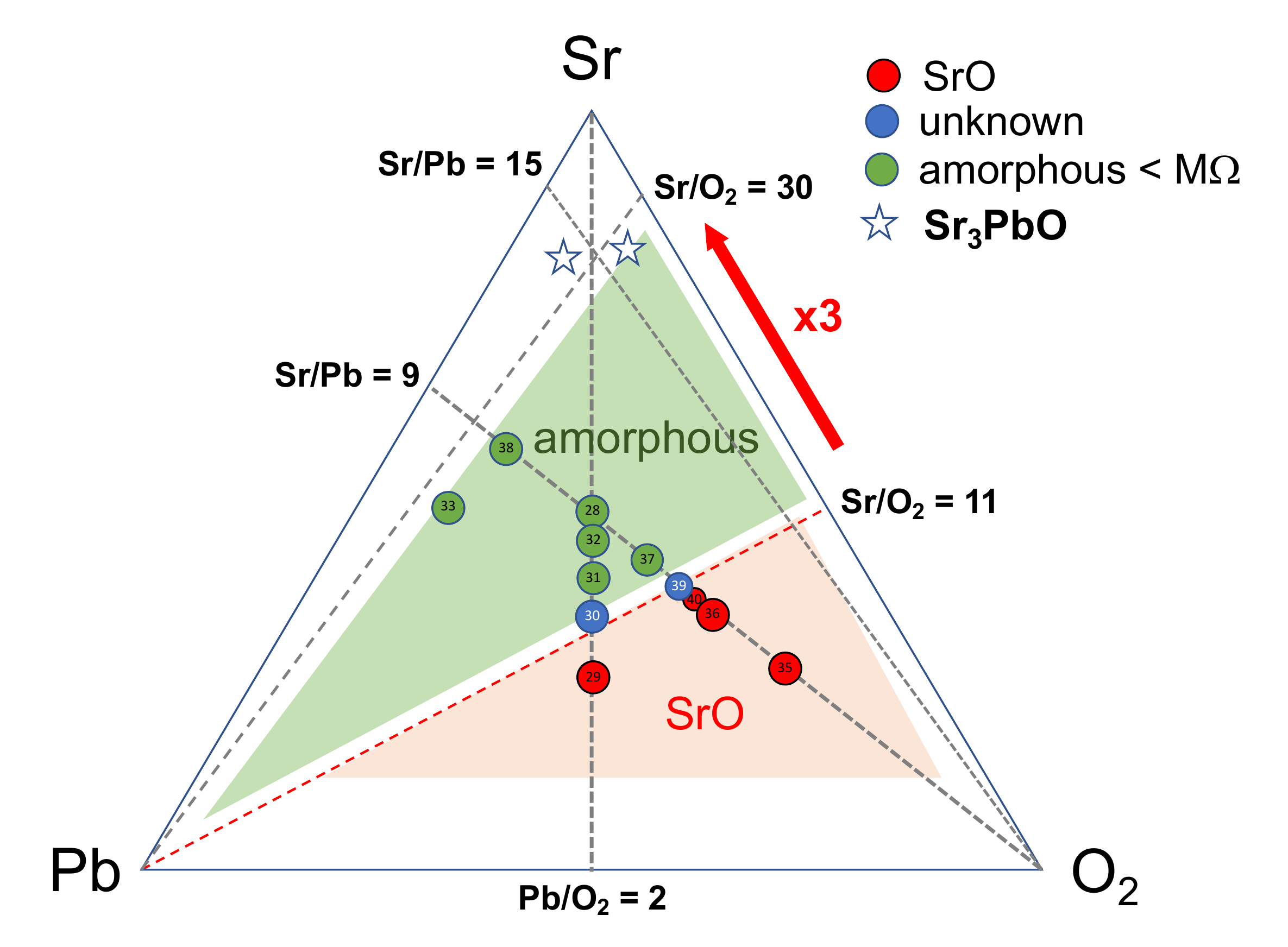}
\caption{%
The ``phase diagram'' of MBE films obtained by XRD as a function of flux ratios (co-deposition). The substrate temperature was 450$^\circ$C. Pb and Sr fluxes were determined by a QCM, while the oxygen flux is reported as a flow rate (sccm); thus, the ratios involving oxygen are not calibrated to atomic ratios. The first two successful films of Sr$_3$PbO are shown as stars. The lesson is that the Sr flux had been too small during the initial phase of the growth efforts. Small numbers inside the circles indicate sample numbers.
}
\label{fig:phase}
\end{figure}

We performed growth using a customized oxide MBE system (Eiko, Japan; Fig. \ref{fig:mbe}). We first review Sr$_3$PbO growth obtained by co-deposition of elemental sources on LAO substrates at a substrate temperature of 450$^\circ$C. No \textit{ex-situ} pre-treatment of the substrates was performed; the as-bought substrates were loaded into the growth chamber and degassed at the growth temperature of the film. As MBE sources, we used pure Sr and Pb from standard effusion cells, and oxygen gas from a mass flow controller. The oxygen gas was later changed to an Ar-O$_2$ mixture. The early attempts of growth at various flux ratios are shown in Fig.~\ref{fig:phase}. Adjusting the flux or determining the appropriate substrate temperature posed a challenge due to the lack of a prior report on MBE growth. This is especially the case when three elements need to be controlled independently, and we do note that other recent MBE works have utilized SnO/SnO$_2$ as a source to circumvent this issue~\cite{Ma_AM_2020,Wu_APL_2021}. To streamline the growth, we used a diagram like the one in Fig.~\ref{fig:phase} to avoid making duplicate films at similar conditions. By this approach, we found that (1) at lower Sr flux, we obtain a SrO phase, largely independent of the Pb flux; (2) upon increasing the Sr flux at a constant O$_2$ and Pb flux, we destroy the SrO phase and obtain an amorphous phase (determined by the lack of any peaks in XRD) with low electrical resistance ($<$M$\Omega$). The phase space for this ``amorphous'' region was large, hampering an immediate success. However, there was a relatively clear boundary where the SrO phase was destroyed by increasing the Sr/O$_2$ ratio. By increasing the Sr flux to reach three times the Sr/O$_2$ value needed to destroy the SrO phase, we obtained the first few films of Sr$_3$PbO. This occurred at impractically large Sr flux of 6 to 7 $\rm{\AA / s}$, as measured by a quartz crystal monitor (QCM), which depleted the Sr cell in 1--2 weeks. Ar-diluted O$_2$ (10\% O$_2$, and later 2\% O$_2$) enabled the use of a lower Sr flux, making the MBE growth more sustainable.

Further optimization with the co-deposition technique was possible by utilizing XRD (00$l$) peaks with odd $l$. In fact, since the (00$l$) peaks with even $l$ originate also from SrO impurity, they do not serve as evidence of an antiperovskite phase. To structurally confirm the antiperovskite phase, (00$l$) peaks with odd $l$ need to be observed. This, in turn, provided an approach to further optimize the growth: it was found that the ratio of the amplitudes between odd and even (00$l$) peaks was maximized at optimized growth conditions~\cite{Samal_APLM_2016}. This could be explained by taking into account long-range structural order. Because the odd (00$l$) peaks in XRD are only allowed due to the distinct layers on the (001) plane, namely, OSr$_2$ and PbSr layers for Sr$_3$PbO, any structural disorder that mixes those layers revives the extinction rule for rock-salt structures that suppresses odd diffraction peaks (as in SrO). For example, an insertion of a SrO layer between two antiperovskite layers could easily impact the long-range structural coherence of the film.

Although experiments using LAO substrates enabled the identification of the growth window, the grown antiperovskite films had limited quality. For Sr$_3$PbO, the best rocking curve of the (001) peak directly grown on LAO showed a full-width half-maximum (FWHM) of 0.88$^\circ$. By using a SrO buffer layer on LAO, we obtained a FWHM = 0.65$^\circ$. By using YSZ as the substrate, the FWHM was further reduced down to 0.43$^\circ$. These results obtained for co-deposition clearly demonstrate the impact of substrates. Substantial improvements in film quality were seen by using lattice-matched YSZ substrates. Next, we show that further improvements in the antiperovskites films could be achieved by utilizing sequential (shuttered) deposition.

\subsection{Sequential Epitaxy}
\label{seq}

After we identified the antiperovskite phase, we tested several strategies to further improve the quality of antiperovskite films grown by MBE. One issue that seemed to impact the quality was the presence of SrO as a dominant impurity phase. Since it is perfectly lattice-matched to antiperovskites, it seemed hard to completely eliminate its formation during standard co-deposition conditions. Therefore, we decided to separate the flux of oxygen from other fluxes by the use of shutters. Also, to better control the O$_2$ at a low flux, a regulated UHV leak valve was installed (instead of a mass flow controller). Both the shutters and the UHV leak valve were controlled by computer via Labview (National Instruments). In most of the sequential epitaxy growths, the Sr flux at the open shutter condition, as monitored by a QCM, was $\sim$0.28--0.30 \AA/s, while the Pb (and later Sn) flux was tuned to achieve a Sr/Pb (or Sr/Sn) ratio of 10--20, which corresponded to 0.015--0.029 \AA/s.

\begin{figure}[h]
\includegraphics[scale=1]{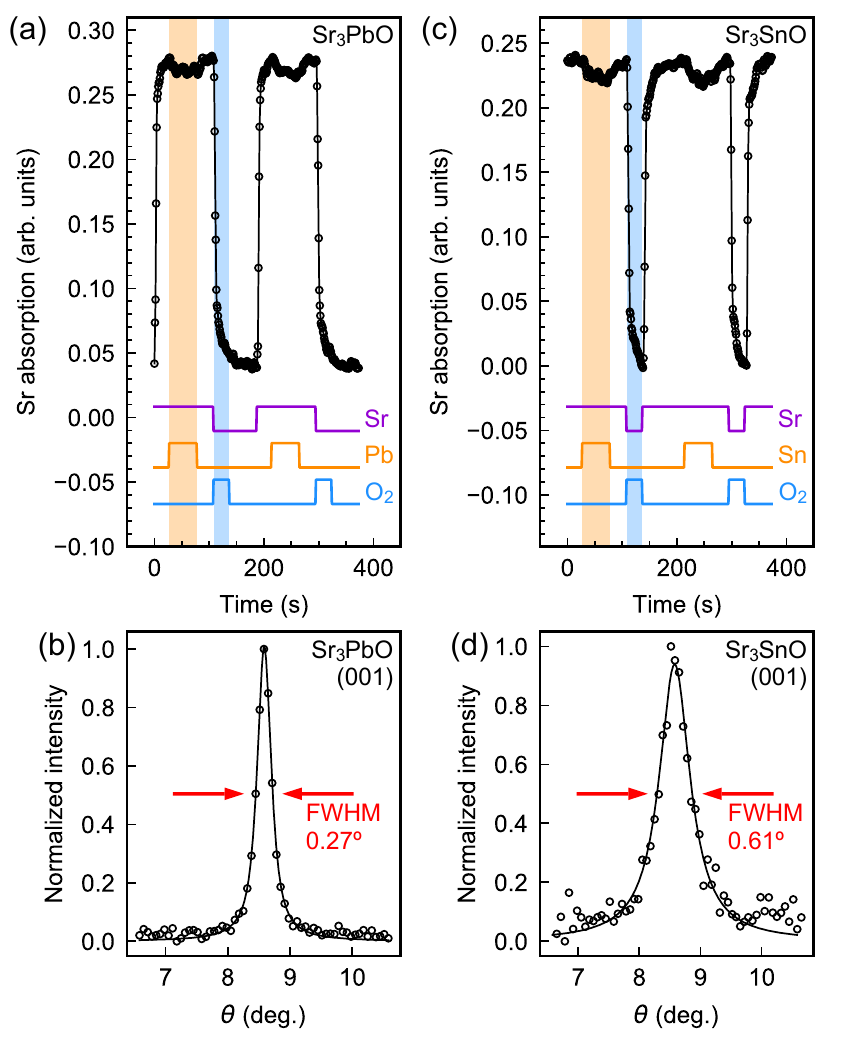}
\caption{%
(a) Sr flux variation during the sequential (element-shuttered) growth of Sr$_3$PbO as monitored by AAS. The absorption signal carries contributions from Sr atoms both impinging the substrate and being re-evaporated from the substrate. The shutter sequences for Sr, Pb, and O$_2$ are also depicted, with the high state corresponding to an open shutter, and the low state corresponding to a closed shutter. (b) XRD rocking curve of the (001) reflection for Sr$_3$PbO (film thickness: 200 nm) obtained using this method. (c) and (d) Corresponding Sr absorption curve, shutter sequences and (001) rocking curve for a Sr$_3$SnO film (film thickness: 300 nm). Both Sr$_3$PbO and Sr$_3$SnO were grown on YSZ substrates at 450$^\circ$C.
}
\label{fig:4}
\end{figure}

The best pattern for sequential growth was identified by trial and error. When all the fluxes were separated, that is, when we deposited Sr, Pb, and O$_2$ one after the other without overlap, the antiperovskite phase did not form. The best result was obtained by opening the Pb shutter in the middle of Sr deposition, while introducing O$_2$ flux at a separate timing [Fig.~\ref{fig:4}(a)]. Atomic absorption spectroscopy (AAS; Accuflux from SVTA) provided insights into the growth mode. In our chamber, AAS was installed via a pair of counter-facing CF16 viewports positioned right below the substrate position. In this position, a large portion of the measured Sr flux was re-evaporated flux from the substrate heating stage, which we confirmed by comparing the Sr flux measured at different substrate temperatures. (Apparently, the Sr flux measured by AAS dropped substantially with decreasing substrate temperature, although the flux independently monitored by a QCM remained constant). This means that the MBE growth is at an adsorption-controlled condition, wherein excess amounts of Sr atoms re-evaporate from the substrate. As shown in Fig.~\ref{fig:4}(a), the Sr flux as seen by AAS exhibited a noticeable decrease when the Pb shutter was opened. This is interpreted as the result of the formation of a PbSr layer on the substrate, preventing the re-evaporation of a portion of Sr that combined with Pb. How the subsequent OSr$_2$ layer is formed upon introducing O$_2$ is less clear, but we speculate that either (i) O$_2$ sticks on the PbSr surface, then is later converted to OSr$_2$ by excess Sr flux, or (ii) the completed PbSr layer flooded with excess Sr atoms (as our sequence does) prepares adsorbed Sr atoms to act as reaction sites with O$_2$. The combination of both processes (i) and (ii) is also possible. The shutter sequence was fine-tuned such that one full cycle corresponded to one unit cell (one PbSr and one OSr$_2$ layer). We confirmed that the growth rate obtained by post-growth thickness measurements (stylus profilometer; Dektak) from several Sr$_3$SnO films (described later) was 0.46 nm/cycle, which is close to the lattice constant of 5.14$\rm{\AA}$ for Sr$_3$SnO. 
    
After the adoption of this sequential growth, we observed an improvement in the film quality, as demonstrated by XRD rocking curves of the (001) peaks, reaching 0.27$^\circ$ for Sr$_3$PbO [Fig.~\ref{fig:4}(b)]. (A further comparison of Sr$_3$PbO films grown via co-deposition and shuttered epitaxy, including carrier density and mobility, is given the Supplemental). With this sequenced approach, we were also able to successfully grow Sr$_3$PbO at substrate temperatures of 370 and 500$^\circ$C, in addition to 450$^\circ$C. At even higher temperatures ($T$ $>$ 500$^\circ$C), the Sr flux needed to be increased substantially to compensate for the increased re-evaporation from the substrate, and thus was judged to be impractical. Growth at temperatures lower than 370$^\circ$C should work as well, although they remain untested.

Having established the sequential growth of Sr$_3$PbO, we began to grow Sr$_3$SnO thin films as well. A key electronic difference between the two arises from the smaller spin-orbit coupling of Sn, which results in a second Dirac point, protected by symmetry against gapping, within 100 meV below $E_F$~\cite{Kariyado_PRM_2017}. We were able stabilize the target antiperovskite phase much faster by adapting the sequential deposition technique, instead of exploring the entire Sr-Sn-O phase diagram with co-deposition. We note some differences in the growths of Sr$_3$SnO and Sr$_3$PbO: First, higher Sr flux was needed in the former. Not only did we increase the Sr/Sn flux ratio ($\sim$12-15 compared to Sr/Pb$\sim$10-12), we also increased the amount of time in each cycle during which the Sr shutter was open [note different sequence in Fig.~\ref{fig:4}(c)]. Second, during the optimization of Sr$_3$SnO films, we uncovered recognizable impurity peaks in XRD in the vicinity of the antiperovskite phase, in contrast to the large ``amorphous'' region near the Sr$_3$PbO antiperovskite phase (Fig.~\ref{fig:phase}). We could achieve a FWHM of 0.61$^{\circ}$ for the (001) reflection of Sr$_3$SnO in a film with no visible impurity peaks in XRD [Fig.~\ref{fig:4}(d)]. Further improvement of the crystallinity and reduction of the FWHM down to 0.35$^{\circ}$ was possible; however, such Sr$_3$SnO films showed an impurity phase in XRD, possibly SrSn (see Supplemental).

\subsection{Solid-source SnO$_2$ for Sr$_3$SnO growth}
\label{alt}

In this section, we describe an alternative MBE growth approach for Sr$_3$SnO films. Instead of controlling Sr, Sn, and O separately, both Ma \textit{et al.} and Wu \textit{et al.} used Sr and the binary oxide SnO$_2$ as solid sources \cite{Ma_AM_2020, Wu_APL_2021}. Upon heating, SnO$_2$ gives rise to SnO and O$_2$ fluxes, which provide Sn and O species presumably close to an ideal atomic ratio of 1:1. Meanwhile, excess amounts of Sr flux are used to achieve adsorption-controlled growth. Confirming this scenario, high Sr/SnO flux ratios of 12.5 \cite{Wu_APL_2021} or 13-17 \cite{Ma_AM_2020} were needed to produce high quality Sr$_3$SnO films.
The growth temperatures were 400-600$^\circ$C \cite{Ma_AM_2020} and 650$^\circ$C \cite{Wu_APL_2021}, showcasing that at least for Sr$_3$SnO, this adsorption-controlled growth can be realized in a wide range of substrate temperatures. This is also reasonable because SnO is much less volatile compared to Sr \cite{Ma_AM_2020}. We note that a much higher Sr flux (1.9--40 \AA/s) was used \cite{Ma_AM_2020} compared to that for the sequential MBE ($\sim$0.3 \AA/s), which likely helped the growth at higher substrate temperatures ($>$500$^{\circ}$C) by overcoming the rapid re-evaporation of Sr from the substrate.

With the MBE-grown Sr$_3$SnO films, Ma \textit{et al.} successfully performed \textit{in-situ} ARPES of the valence bands of Sr$_3$SnO, which showed good agreement with DFT calculations~\cite{Ma_AM_2020}. Wu \textit{et al.} used scanning electron microscopy (SEM) to investigate their MBE-grown Sr$_3$SnO films\cite{Wu_APL_2021}. In Sn-rich films that exhibited similar transport properties to that of pure stoichiometric films and showed no impurity peaks in XRD, SEM nevertheless revealed the presence of impurity phases as dark columnar regions. Hence, the use of microscopy and other advanced characterization tools may provide a fuller picture of antiperovskite films.

\subsection{Transport}
\label{trans}
The antiperovskite films suffer from extreme air sensitivity and degrade within seconds of exposure to ambient environment, as evidenced by a pronounced change in color from metallic silver to opaque brown. To characterize the films after MBE growth, two pathways are possible. (1) For \textit{ex-situ} measurements, we transported the films in a vacuum suitcase from the load lock of the MBE to a glove box filled with inert Ar gas (inset of Fig.~\ref{fig:mbe}). Inside the glove box, we evaporated a Au capping layer, which then enabled the antiperovskite films to survive in air for several hours during XRD measurements or thickness measurements. For transport measurements of the films, we deposited Au electrodes inside the glove box, covered the remaining areas of the film with a grease droplet (Apiezon N), and performed wiring onto a puck either with Ag epoxy (EPOTEK E4110; room-temperature cure) inside the glove box, or bonding outside the glove box. These steps allowed the films to survive when transferring the wired device to a PPMS. (2) For measurements in UHV, these were performed either in the MBE growth chamber for RHEED, or in an adjoining chamber for STM (Fig.~\ref{fig:mbe}), or transported via a vacuum suitcase to a separate chamber for XPS and LEED.   

\begin{figure}[!tb]
\includegraphics[scale=1]{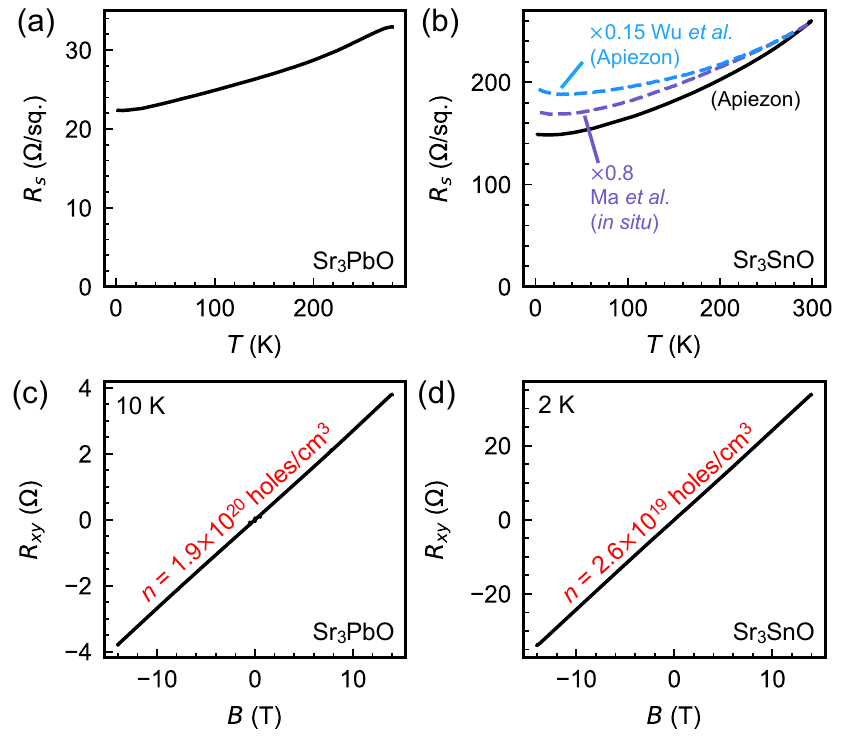}
\caption{Transport characterization of antiperovskite films. (a) and (b) Temperature-dependent sheet resistances of a Sr$_3$PbO (120 nm thick) and Sr$_3$SnO (100 nm thick) film, respectively. In (b), similar resistance curves of Sr$_3$SnO films from Refs.~\cite{Ma_AM_2020, Wu_APL_2021} are scaled and overlaid, showing qualitative consistency among films measured \textit{in situ}, or \textit{ex situ} with a protection layer of Apiezon grease. (c) and (d) Transverse resistances in a perpendicular magnetic field. The carrier densities and mobilities are 1.9 $\times$ 10$^{20}$ holes/cm$^{3}$ and 121 cm$^2$/(Vs)$^{-1}$ at 10 K for the Sr$_3$PbO film and 2.6 $\times$ 10$^{19}$ holes/cm$^{3}$ and 162 cm$^2$/(Vs)$^{-1}$ at 2 K for the Sr$_3$SnO film.
}
\label{fig:trans}
\end{figure}

Both Sr$_3$PbO and Sr$_3$SnO revealed metallic behavior with similar residual resistivity ratios around 1.5--2 [Figs.~\ref{fig:trans}(a) and \ref{fig:trans}(b)]. The resistance curve of Sr$_3$SnO is qualitatively similar to that of Refs.~\cite{Ma_AM_2020, Wu_APL_2021} [overlaid in Fig.~\ref{fig:trans}(b)], showing good consistency among the MBE-grown Sr$_3$SnO films of different groups, measured both with an \textit{in-situ} four-probe setup~\cite{Ma_AM_2020} and \textit{ex-situ} with a grease protection layer (here and \cite{Wu_APL_2021}). Hall measurements [Figs.~\ref{fig:trans}(c) and \ref{fig:trans}(d)] point to hole carriers, presumably arising from Sr vacancies. While the hole carrier concentration could be tuned by varying the Sr/Pb and Sr/Sn ratios, the lowest concentrations we could attain were around 1$\times$10$^{20}$ cm$^{-3}$ for Sr$_3$PbO and 1$\times$10$^{19}$ cm$^{-3}$ for Sr$_3$SnO. For Sr$_3$SnO, the corresponding Fermi energy is within the range of the Dirac dispersion, making the films ideal for magnetotransport measurements~\cite{Nakamura_NatComm_2020, Huang_PRR_2021, Wu_APL_2021}.

\subsection{Surface characterization}
\label{surf}

We now direct our focus to the characterization of antiperovskite surfaces. It is well established that the perovskite oxides exhibit rich surface chemistry. For example, the workhorse substrate SrTiO$_3$ hosts a plethora of surface reconstructions, including (2$\times$1), (2$\times$2), $c$(4$\times$2), $c$(4$\times$4), $c$(4$\times$6), (6$\times$2), ($\sqrt{5}$$\times$$\sqrt{5}$)$R$26.6$^{\circ}$, and $\sqrt{13}$$\times$$\sqrt{13}$)$R$33.7$^{\circ}$ superstructures for the nonpolar (001) surface alone~\cite{Erdman_JACS_2003}. Many of these reconstructions are oxygen deficient and are formed simply by annealing at high temperatures in UHV. Different reconstructions also appear during the MBE growth of SrTiO$_3$ films by making fine changes in flux ratios~\cite{Kajdos_APL_2014}. In contrast, there is scant information regarding surface reconstructions in antiperovskite oxides. As previously described, the Sr$_3$PbO(001) and Sr$_3$SnO(001) surfaces are polar (type III according to Tasker's classification~\cite{Tasker_JPC_1979}), and without an atomic or electronic reconstruction, the electrostatic potential would diverge at the surface. It is also natural to ask whether Sr vacancies in Sr$_3$PbO and Sr$_3$SnO can play an analogous role to O vacancies in perovskites in stabilizing surface reconstructions. Understanding the surface atomic structure of antiperovskites is crucial, since topological~\cite{Chiu_PRB_2017} and/or spin-polarized~\cite{Arras_PRB_2021} surface states have been predicted, and their properties would be affected by reconstructions. Edge states arising from higher-order topology are also predicted~\cite{Fang_PRB_2020}, so investigating terrace structures at surfaces would also be useful. Many systematic studies are needed in the antiperovskites; here, we provide a brief overview of preliminary investigations in this direction.

XPS is a surface-sensitive technique that probes the core levels of elements and reveals their valence state and chemical environment~\cite{Moulder_1992}. By measuring at normal and grazing emission angles, the probing depth can be varied. In the case of Sr$_3$PbO and Sr$_3$SnO, it is instructive to check whether anionic Pb and Sn can be stabilized in thin films, given that the massive Dirac cones are formed partly by the nearly filled $p$ shells of Pb and Sn. (Although the formal oxidation states of Pb and Sn are $-$4, the Bader charges computed by DFT is closer to $-$2, indicating a degree of covalency~\cite{Huang_PRM_2019}). We found that indeed in our films, the Pb 4$f$ and Sn 3$d$ core levels showed a component with lower binding energy than metallic Pb and Sn, consistent with anionic Pb and Sn~\cite{Huang_PRM_2019}. However, traces of cationic and/or neutral Pb and Sn were also observed within a surface layer, estimated to be on the order of 1 nm thick. This surface layer appeared to be stronger in Sr$_3$SnO. Furthermore, we observed both cationic and neutral Sn on the surface of Sr$_3$SnO, but only cationic Pb on the surface of Sr$_3$PbO.   

\begin{figure}[t]
\includegraphics[scale=1]{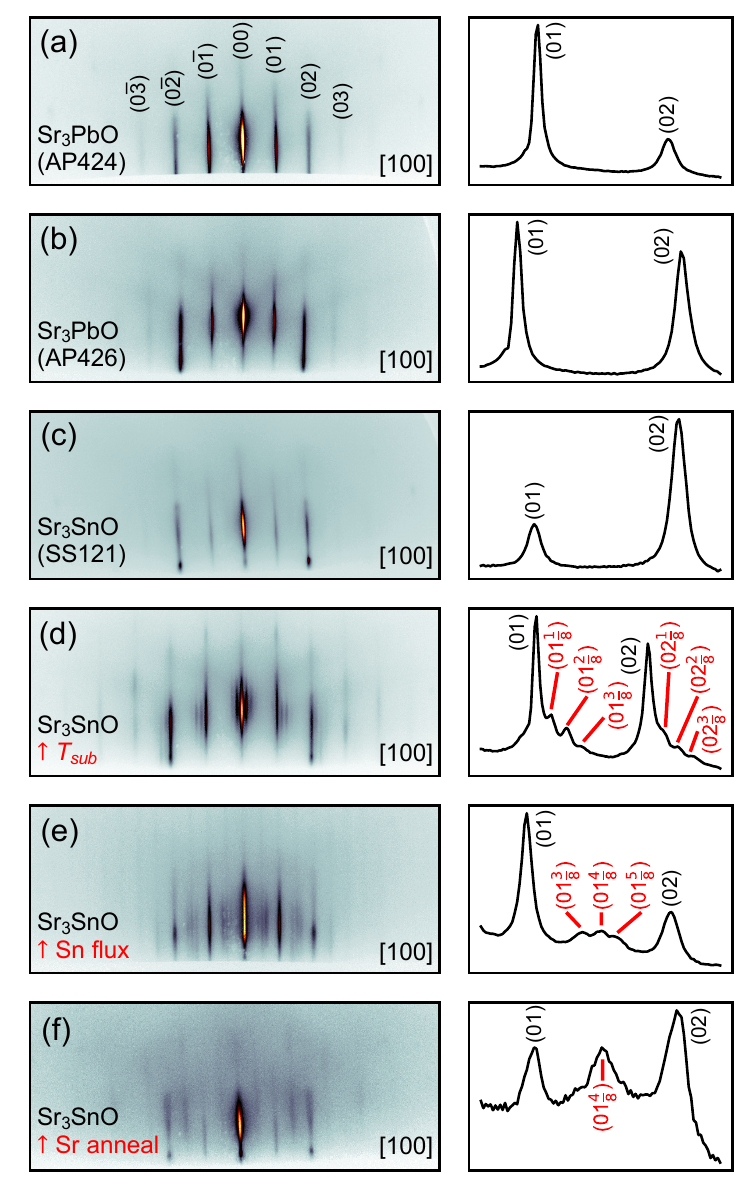}
\caption{RHEED images of various (a) and (b) Sr$_3$PbO(001) and (c)--(f) Sr$_3$SnO(001) films taken along the [100] direction. Electron energy: 15 keV. Corresponding line cuts intersecting the (01) and (02) streaks are shown. Sample numbers correspond to those in Fig.~\ref{fig:6}. Superstructure streaks in Sr$_3$SnO(001) films were obtained by tuning growth conditions, such as (d) increasing the substrate temperature by 50$^{\circ}$C, (e) increasing the Sn flux, or (f) annealing the film in Sr flux for several hours after growth.}
\label{fig:5}
\end{figure}

Further insights into the surfaces of Sr$_3$PbO and Sr$_3$PbO could be derived from LEED and RHEED measurements. Our LEED measurements revealed 1$\times$1 diffraction spots consistent with the expected (001) surfaces of Sr$_3$PbO and Sr$_3$SnO~\cite{Huang_PRM_2019}. RHEED measurements similarly exhibited streaks indicative of a flat, 2D film [Fig.~\ref{fig:5}(a)-(c)]. We note that measurements of RHEED during the film deposition were rather useful, because we could differentiate between rock-salt SrO and antiperovskite by the appearance of (0$l$) streaks with odd $l$ that are extinct in the former, but not in the latter (see supplemental). In Sr$_3$SnO, we found that it was possible to induce superstructure streaks in the RHEED pattern by increasing the substrate temperature or tuning the Sr/Sn flux ratio from optimized values. Three examples are shown in Figs.~\ref{fig:5}(d)--(f). Despite differences in their patterns, the superstructure streaks appear to be multiples of $l$ = 1/8, which suggests that the underlying real-space pattern has a periodicity of 8$a$. In Fig.~\ref{fig:5}(d), we observe $l$ = 1/8, 2/8, and 3/8 superstructure streaks; in Fig.~\ref{fig:5}(e), $l$ = 3/8, 4/8, and 5/8 streaks; in Fig.~\ref{fig:5}(f), $l$ = 4/8 streaks. These patterns are reminiscent of the $4a$, $5a$, and $6a$ superstructure streaks detected in La$_{2-x}$Sr$_x$CuO$_4$ thin films and associated with ordering of oxygen vacancies~\cite{Suyolcu_JVSTA_2022, Wu_SR_2017}. In the case of Sr$_3$SnO, future systematic studies are needed to map out the full phase diagram of surface reconstructions. In the case of Sr$_3$PbO, despite tuning the substrate temperature and Sr/Pb flux ratio, we consistently observed only a 1$\times$1 surface with no superstructure.   

\begin{figure*}[!t]
\includegraphics[width=\textwidth,clip]{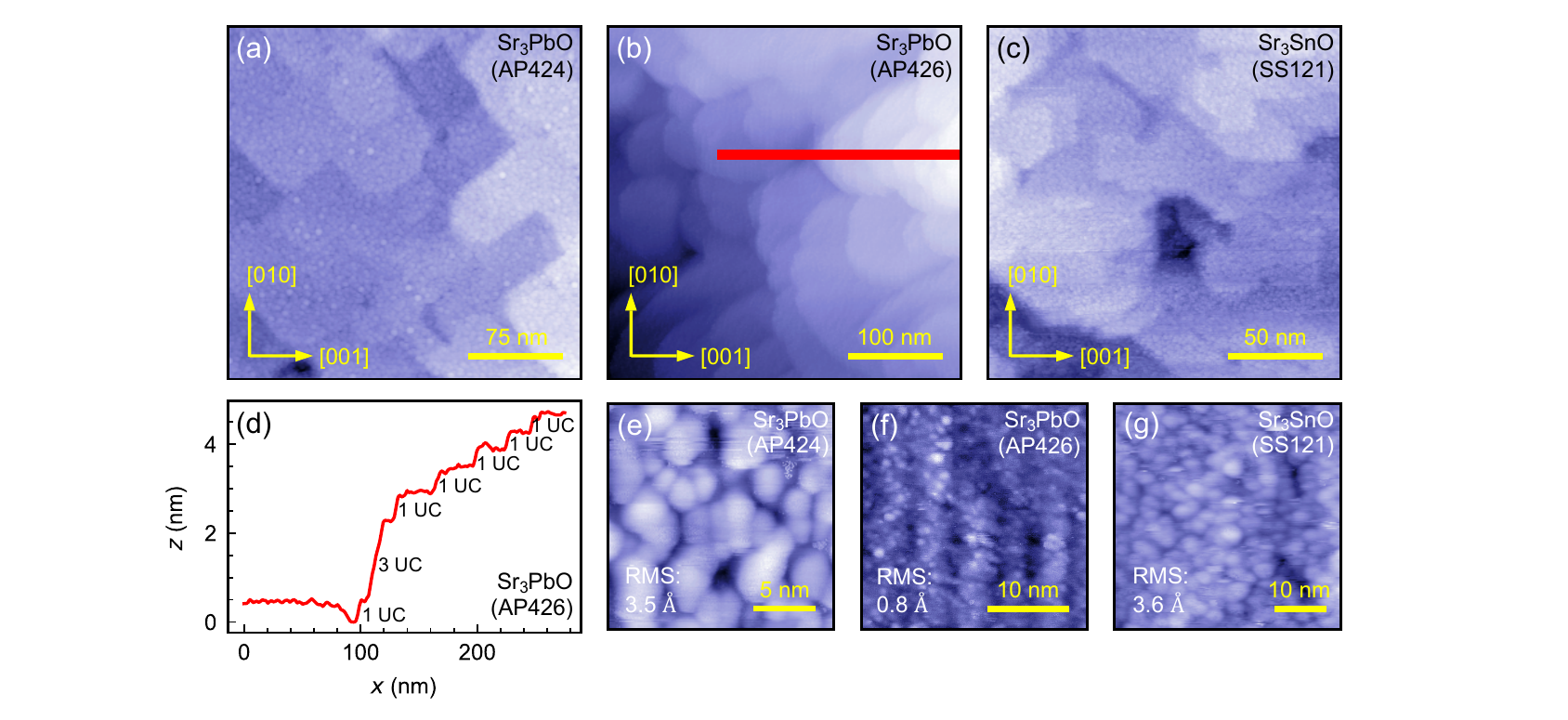}
\caption{STM topographic images of (a) and (b) Sr$_3$PbO and (c) Sr$_3$SnO thin films, revealing step edges with preferential orientation along [110] and [1$\bar{1}$0]. Setpoints: (a) $-$1 V, 10 pA, (b) $-$2 V, 5 pA, (c) $-$1.75 V, 10 pA. (d) Averaged line cut along the horizontal red line in (b). The steps mostly correspond to one unit cell (UC). (e)--(g) Scans over small areas on a single terrace, revealing different kinds of globular surface morphologies. The root-mean-square (RMS) roughness are 3.5, 0.8, and 3.6~\AA, respectively. Setpoints: (e) $-$1 V, 10 pA, (f) $-$1 V, 5 pA,  (g) $-$1.5 V, 10 pA.  
}
\label{fig:6}
\end{figure*}

We performed room-temperature STM (Unisoku, Japan) on several antiperovskite films immediately after MBE growth. Generally, we observed terraces on the order of tens of nanometers in lateral size [Fig.~\ref{fig:6}(a)--(c)]. The step edges appear to be preferentially oriented along the [110] and [1$\bar{1}$0] directions, especially in Fig.~\ref{fig:6}(a), but steps aligned along [100] and [010] were also observed in Fig.~\ref{fig:6}(c), as well as rounder steps that are more densely bunched [Fig.~\ref{fig:6}(b)]. A sample line cut in Fig.~\ref{fig:6}(d) shows that the step heights are consistent with integer multiples of the antiperovskite unit cell.

Our STM measurements failed to resolve any atomic lattices. Instead, we observed different surface morphologies, consisting of globules of varying sizes [Figs.~\ref{fig:6}(e)--\ref{fig:6}(g)], which sometimes showed quasi-ordering into stripes [Fig.~\ref{fig:6}(f)]. The STM images showed little correspondence with RHEED patterns taken on the same films. For example, the Sr$_3$PbO films AP424 and AP426 have the same RHEED pattern with 1$\times$1 streaks [Figs.~\ref{fig:5}(a) and \ref{fig:5}(b)], but STM did not resolve a 1$\times$1 atomic lattice in either case [Figs.~\ref{fig:6}(e) and \ref{fig:6}(f)], and the terracing behavior is quite distinct. In the case of Sr$_3$SnO, we were similarly unable to resolve a 1$\times$1 atomic lattice [sample SS121; Fig.~\ref{fig:6}(g)], even when RHEED patterns indicated such a structure [Fig.~\ref{fig:5}(c)]. 

This discrepancy between RHEED and STM has previously been reported in the perovskite SrTiO$_3$(001). Although simultaneous RHEED measurements showed sharp streaks associated with a 1$\times$1 surface, STM revealed a disordered, globular surface~\cite{Iwaya_STAM_2018}. One hypothesis is that an amorphous layer that is invisible to electron diffraction covers the 1$\times$1 surface. In our case, this could be the $\sim$1-nm-thick surface layer detected in XPS. In La$_{1-x}$Ca$_x$MnO$_3$ thin films, a similar phenomenology has been attributed to the result of an unstable growth front~\cite{Tselev_ACSNano_2015}, which may be more significant for polar surfaces. As the constituent atoms land on the surface of the film, they cluster in highly disordered, 3D globules, for which STM cannot resolve an atomic lattice. As more atoms are deposited and these globules become buried beneath the surface, their crystallinity improves, although defects and vacancies get incorporated. Hence, the surface layer is always disordered, whereas the layers beneath are crystalline. This mechanism may explain the appearance of clear step edges and terrace structure in Sr$_3$PbO and Sr$_3$SnO, indicating epitaxial growth, yet the appearance of globules on each terrace. In the case of SrTiO$_3$(001), a 1$\times$1 surface could be visualized not with STM, but with non-contact AFM~\cite{Sokolovic_PRM_2019}. Furthermore, once the SrTiO$_3$(001) surface reconstructs and deviates from 1$\times$1, STM can resolve the different and diverse reconstructions. In our case, we could not obtain stable STM imaging on Sr$_3$SnO(001) films showing superstructure streaks in RHEED, nor resolve the superstructures.      

In the characterization of MBE-grown antiperovskite films, RHEED is more sensitive than XRD, but STM is even more sensitive than RHEED. RHEED can resolve the appearance of surface reconstructions in Sr$_3$SnO(001) films that otherwise show the same XRD patterns. However, STM can resolve different terrace structures and surface morphologies in Sr$_3$PbO(001) films that otherwise show the same RHEED patterns. Our preliminary measurements indicate that antiperovskites show a similar richness in surface chemistry that is well explored in the perovskite oxides, but many more systematic studies are needed. 


\section{Conclusion}

We have demonstrated pathways for the MBE growth of the antiperovskites Sr$_3$PbO(001) and Sr$_3$SnO(001). By using sequential (element-shuttered) MBE that separates oxygen flux from other fluxes (Sr, Pb, or Sn), high quality films with sharp XRD rocking curves were obtained at relatively low substrate temperatures of 450$^\circ$C. Complex surface reconstructions that indicate a long-range periodicity of eight times the lattice constant were observed in Sr$_3$SnO(001) by deviating from an optimal growth condition, whereas signs of reconstruction were absent in Sr$_3$PbO(001). Such long-range patterns are reminiscent of the ordering of oxygen vacancies in La$_{2-x}$Sr$_x$CuO$_4$ films, but the microscopic origin in the case of the antiperovskites needs to be clarified in subsequent studies. Looking towards the future, there is still much to be improved in terms of film quality, especially in the light of the ultrahigh crystallinity and carrier mobilities that have been recently achieved in the MBE of conventional oxides~\cite{schlom_2008_JACeram,son2010_NatMater,falson2016mgzno}. Further advancement in the control of material quality will most certainly open a novel research arena at the intersection of correlated electrons and topological electronic states. We also anticipate that novel electronic and magnetic phases at surfaces or heterointerfaces of antiperovskites are within experimental reach by using MBE. The exploration of such avenues will likely involve the seamless integration of advanced experimental apparatuses with the MBE system, even beyond what is outlined in this Perspective.

\section{Acknowledgement}

We thank D. Samal and J. Merz for contributions in the development of antiperovskite MBE. We are also grateful to K. Pflaum, M. Dueller, U. Engelhardt, and C. M{\"u}hle for technical support. D.H. acknowledges support from a Humboldt Research Fellowship for Postdoctoral Researchers.

\section{Supplementary Material}

See supplementary material for additional XRD, RHEED, and transport data of antiperovskites grown by MBE.

\bibliography{Main}

\end{document}


\onecolumngrid

\setcounter{figure}{0}
\setcounter{equation}{0}
\setcounter{table}{0}
\setcounter{section}{0}
\setcounter{subsection}{0}
\makeatletter
\renewcommand{\thefigure}{S\@arabic\c@figure}
\renewcommand{\theequation}{S\@arabic\c@equation}
\renewcommand{\thetable}{S\@arabic\c@table}
\newcounter{SIfig}
\renewcommand{\theSIfig}{S\arabic{SIfig}}
\newcounter{SIeq}
\renewcommand{\theSIeq}{S\arabic{SIeq}}

\section*{Supplemental}

\subsection*{Supplementary Note 1: X-ray Diffraction (XRD)}

\begin{figure}[h]
\includegraphics[scale=1]{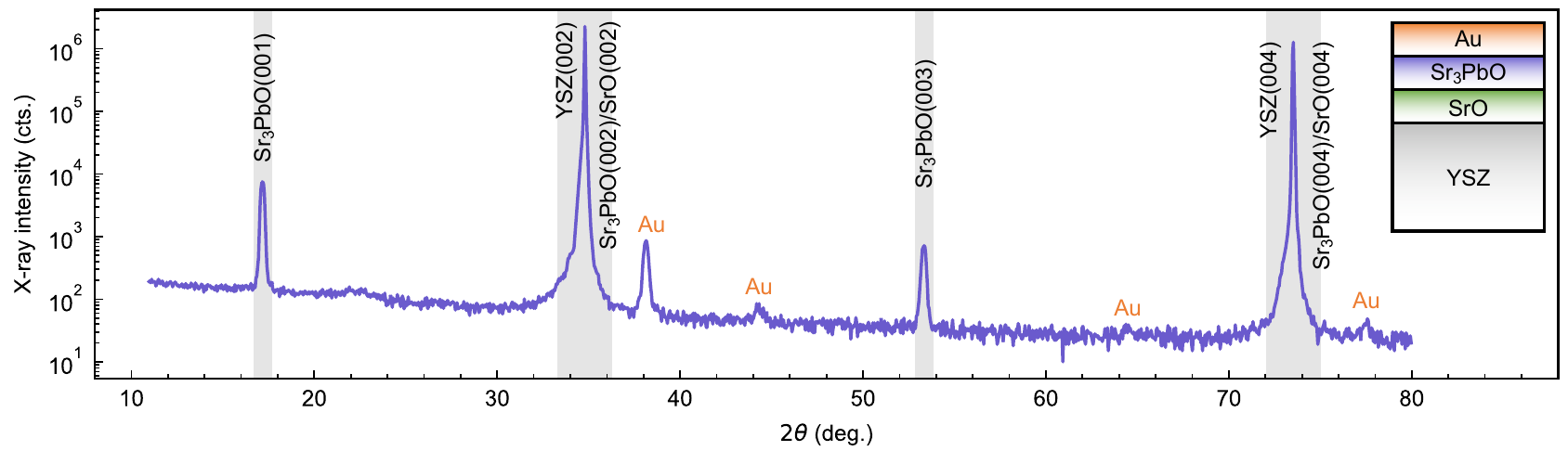}
\caption{XRD $\theta$-2$\theta$ scan of the Sr$_3$PbO film shown in Figs.~4(a) and 4(b) of the main text, taken with a Cu $K_{\alpha}$ source. The inset schematic shows the components of the sample, including a YSZ substrate, SrO buffer layer, Sr$_3$PbO film, and Au capping layer. Their corresponding Bragg reflections are labeled.}
\refstepcounter{SIfig}
\label{FigXRD_AP}
\end{figure}

\begin{figure}[h]
\includegraphics[scale=1]{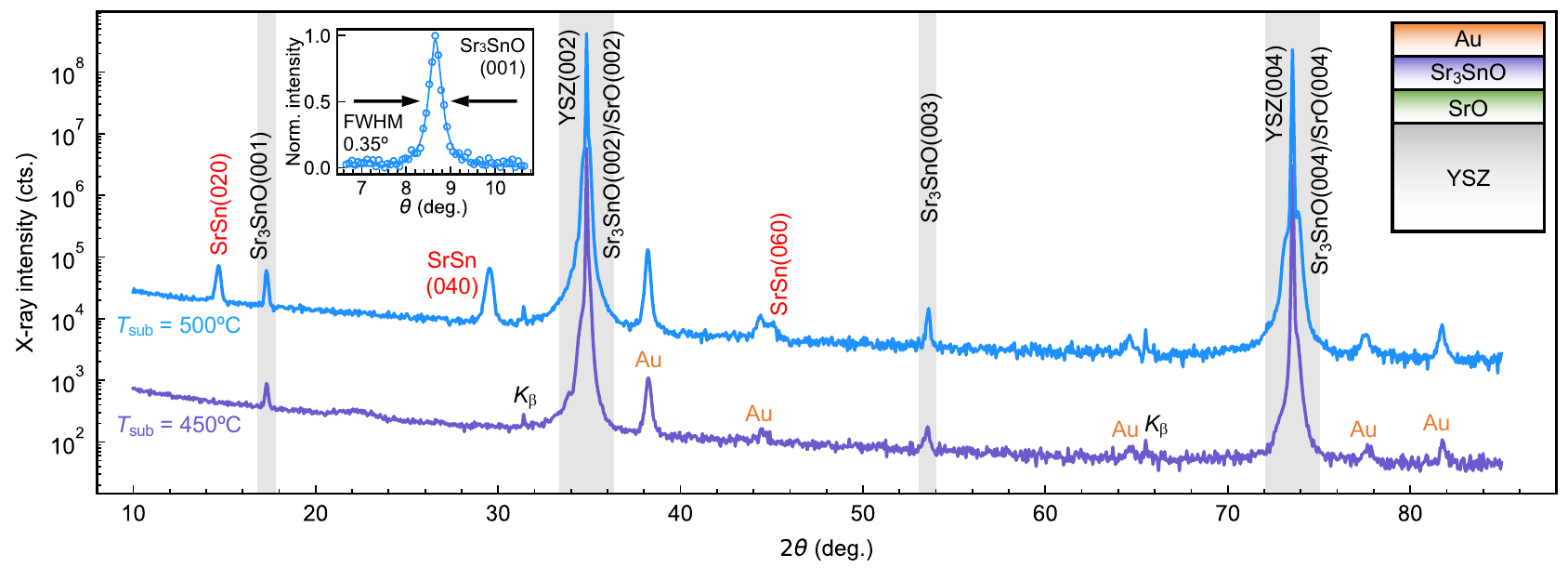}
\caption{XRD $\theta$-2$\theta$ scan of the Sr$_3$SnO film shown in Figs.~4(c) and 4(d) of the main text, grown at a substrate temperature $T_{\textrm{sub}}$ = 450$^{\circ}$C, as well as an additional film grown at 500$^{\circ}$C. The inset schematics show the components of the samples, as well as the (001) rocking curve of the film grown at 500$^{\circ}$C. There are spurious reflections of the dominant substrate peaks, YSZ(002) and YSZ(004), due to a small unfiltered Cu K$_{\beta}$ component.}
\refstepcounter{SIfig}
\label{FigXRD}
\end{figure}

Figures \ref{FigXRD_AP} and \ref{FigXRD} show the XRD $\theta$-2$\theta$ scans for Sr$_3$PbO and Sr$_3$SnO films. In the case of Sr$_3$SnO, when the substrate temperature was increased from 450 to 500$^{\circ}$C, the full-width half-maximum (FWHM) of the rocking curve for the (001) peak was reduced from 0.61$^{\circ}$ (Fig.~4(d); main text) to 0.35$^{\circ}$ (inset of Fig.~\ref{FigXRD}). However, additional peaks corresponding to an impurity phase appeared. The 2$\theta$ values of the three additional peaks are consistent with the (020), (040), and (060) reflections of orthorhombic SrSn, which was also detected as a by-product in the synthesis of single crystals of the layered antiperovskite-type nitride Sr$_7$N$_2$Sn$_3$ \cite{Yamane_ZNB_2021}. We note that Ma \textit{et al.}~\cite{Ma_AM_2020} also detected a SrSn impurity phase in Sr$_3$SnO films grown at an elevated temperature above 600$^{\circ}$C, but with additional orientations, including the (130), (022), and (132) planes.    

\subsection*{Supplementary Note 2: Reflection high-energy electron diffraction (RHEED) during film deposition}

\begin{figure}[h]
\includegraphics[scale=1]{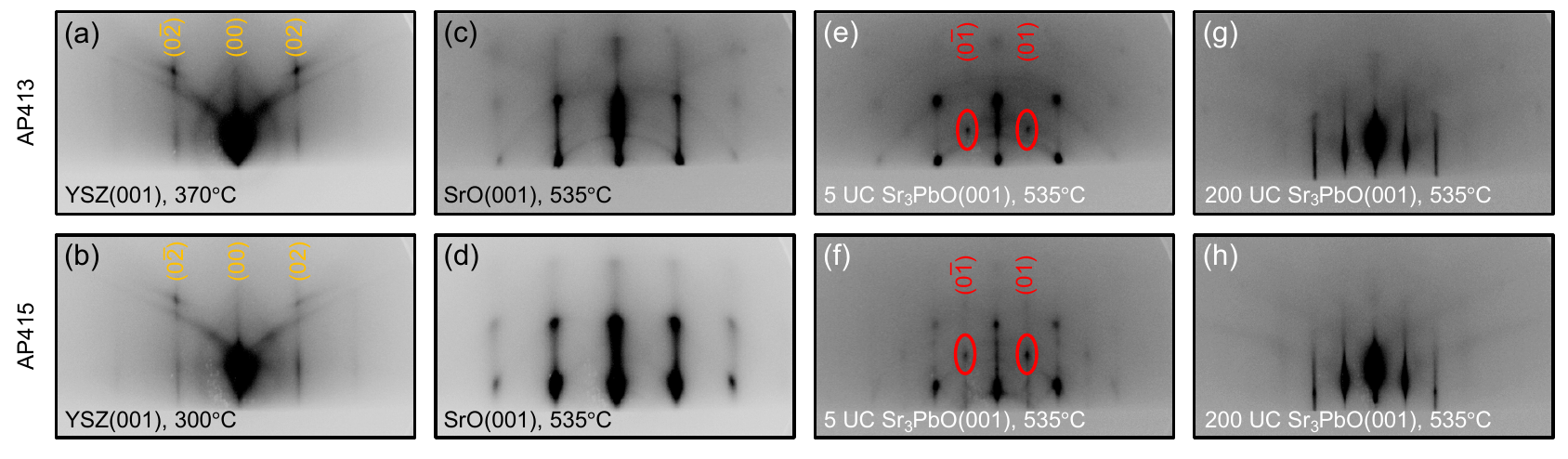}
\caption{RHEED images of two Sr$_3$PbO films (samples AP413 and AP415) at various stages of growth: (a) and (b) pre-deposition, (c) and (d) deposition of SrO buffer layer, (e) and (f) deposition of five unit cells (UCs) of Sr$_3$PbO, and (g) and (h) deposition of 200 UC Sr$_3$PbO. The images were all acquired along the [001] direction. Electron energy: 15 keV.}
\refstepcounter{SIfig}
\label{FigSRHEED}
\end{figure}

Figure \ref{FigSRHEED} presents additional RHEED images taken at intermediate stages of the Sr$_3$PbO growth (shuttered epitaxy). Even during the deposition of the first few unit cells of Sr$_3$PbO, the RHEED pattern can be differentiated from that of YSZ or SrO by the appearance of (0$l$) diffraction with odd $l$ [red circles; Figs.~\ref{FigSRHEED}(e) and \ref{FigSRHEED}(f)]. At five unit cells, the film appears to be 3D, as evidenced by the spotty patterns [Figs.~\ref{FigSRHEED}(c) and \ref{FigSRHEED}(d)]. This 3D morphology may be inherited from the 3D SrO buffer layer with similar spotty patterns. However, after more layers, the film becomes 2D, as evidenced by the streaky patterns [Figs.~\ref{FigSRHEED}(g) and \ref{FigSRHEED}(h)].  

\subsection*{Supplementary Note 3: Comparison of Sr$_3$PbO films grown via co-deposition and shuttered epitaxy}

\begin{figure}[h]
\includegraphics[scale=1]{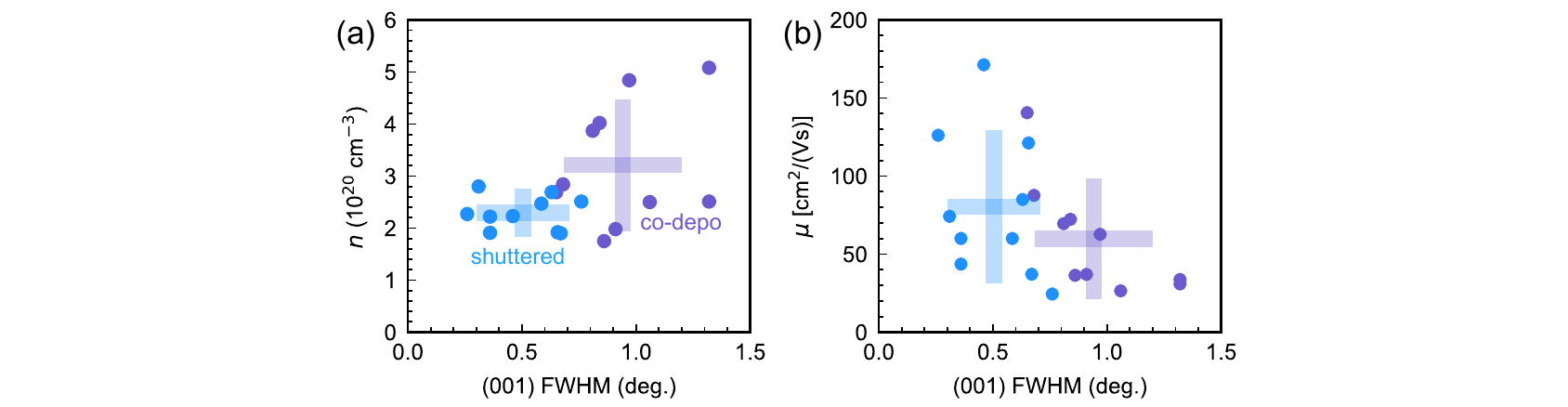}
\caption{Scatter plots comparing ten Sr$_3$PbO films grown via co-deposition and ten grown via shuttered epitaxy, all on YSZ substrates with 450$^{\circ}$C substrate temperature. (a) Hole carrier density ($n$) vs. FWHM of the Sr$_3$PbO(001) Bragg reflection in XRD. (b) Mobility ($\mu$) vs. FWHM. The shaded bars mark one standard deviation above and below the mean values.}
\refstepcounter{SIfig}
\label{FigStats}
\end{figure}

Figure \ref{FigStats} compares the FWHM of the Sr$_3$PbO(001) peak, hole carrier density, and mobility of ten films grown via co-deposition (II.A. of main text), and ten films grown via shuttered epitaxy (II.B. of main text). All films were deposited on YSZ substrates held at 450$^{\circ}$C. We observed a reduction in the FWHM of the (001) peak from 0.94$\pm$0.22$^{\circ}$ for the co-deposition samples to 0.51$\pm$0.17$^{\circ}$ for the shuttered epitaxy samples. The corresponding carrier densities and mobilities are (3.2$\pm$1.1)$\times$10$^{20}$ cm$^{-3}$ and 60$\pm$34 cm$^{2}$/(Vs) for the co-deposition samples and (2.3$\pm$0.3)$\times$10$^{20}$ cm$^{-3}$ and 80$\pm$44 cm$^{2}$/(Vs) for the shuttered epitaxy samples. 
 
\bibliography{Suppl}